\documentclass[aps,prl,superscriptaddress,11pt]{revtex4-2}

\usepackage{color}
\usepackage{xcolor}
\usepackage{graphicx}
\usepackage{amsmath}
\usepackage{amsfonts}
\usepackage{soul}
\usepackage[normalem]{ulem}
\usepackage{lipsum}
\usepackage{bibentry}
\usepackage{float}

\usepackage{silence}
\begin{document}

\title{Limits of nonlinear and dispersive fiber propagation for \\ an optical fiber-based extreme learning machine}

\author{Andrei V. Ermolaev}
\affiliation{Universit\'{e} Marie et Louis Pasteur, Institut FEMTO-ST, CNRS UMR 6174, 25000 Besan\c{c}on, France}

\author{Mathilde Hary}
\affiliation{Photonics Laboratory, Tampere University, FI-33104 Tampere, Finland}
\affiliation{Universit\'{e} Marie et Louis Pasteur, Institut FEMTO-ST, CNRS UMR 6174, 25000 Besan\c{c}on, France}

\author{Lev Leybov}
\affiliation{Photonics Laboratory, Tampere University, FI-33104 Tampere, Finland}

\author{Piotr Ryczkowski}
\affiliation{Photonics Laboratory, Tampere University, FI-33104 Tampere, Finland}

\author{Anas Skalli}
\affiliation{Universit\'{e} Marie et Louis Pasteur, Institut FEMTO-ST, CNRS UMR 6174, 25000 Besan\c{c}on, France}

\author{Daniel Brunner}
\affiliation{Universit\'{e} Marie et Louis Pasteur, Institut FEMTO-ST, CNRS UMR 6174, 25000 Besan\c{c}on, France}

\author{Go\"{e}ry Genty}
\affiliation{Photonics Laboratory, Tampere University, FI-33104 Tampere, Finland}

\author{John M. Dudley}
\affiliation{Universit\'{e} Marie et Louis Pasteur, Institut FEMTO-ST, CNRS UMR 6174, 25000 Besan\c{c}on, France}
\affiliation{Institut Universitaire de France, Paris, France}

\begin{abstract}
We report a generalized nonlinear Schr\"{o}dinger equation simulation model of an extreme learning machine (ELM) based on optical fiber propagation. Using the MNIST handwritten digit dataset as a benchmark, we study how accuracy depends on propagation dynamics, as well as parameters governing spectral encoding, readout, and noise. For this dataset and with quantum noise limited input, test accuracies of : over 91\% and 93\% are found for propagation in the anomalous and normal dispersion regimes respectively. Our results also suggest that quantum noise on the input pulses introduces an intrinsic penalty to  ELM performance.  
\end{abstract}

\maketitle

\clearpage

There is currently intense interest in developing photonic-based artificial intelligence hardware. Often described as optical neuromorphic computing \cite{Psaltis-1990,Wetzstein-2020}, experiments have reported photonic neural networks \cite{Miscuglio-2018, Zuo-2019, Jha-2020,Liu-2021,Oguz-2024,Dinc-2024}, reservoir computers \cite{Larger-2012,Duport-2012,Yildirim-2023}, and extreme learning machines (ELMs) \cite{Ortin-2015,Tegin-2021,Oguz-2024}. One specific focus has been applying nonlinear wave propagation directly as a computational resource \cite{Marcucci-2020}, and recent experiments have reported ELM classification based on nonlinear propagation and supercontinuum generation in optical fiber \cite{Fischer-2023,Lee-2024,Hary-2025,Saeed-2025}. However, whilst fiber-based ELM performance has been studied experimentally over broad parameter ranges \cite{Hary-2025, Saeed-2025}, guidance from a robust simulation model is clearly needed in order to gain a more complete picture of such nonlinear wave-based computing. 

Here we report such an end-to-end numerical model of nonlinear fiber propagation in an ELM architecture, where we analyze handwritten digit classification using the MNIST Digit dataset of 60,000 training and 10,000 test images. (The Supplemental document shows results for the more complex MNIST Fashion dataset.) We specifically study how the ELM-based classification accuracy depends on data encoding, propagation dynamics and readout, as well as input noise. Although some parameter dependencies have been studied experimentally \cite{Hary-2025, Saeed-2025, Muda-2025}, simulations provide significant new insights in evaluating the effect of quantum and technical noise sources, and in exploring different dynamical regimes.  

\begin{figure}[hbt]
\centering
\includegraphics[width=0.75\linewidth]{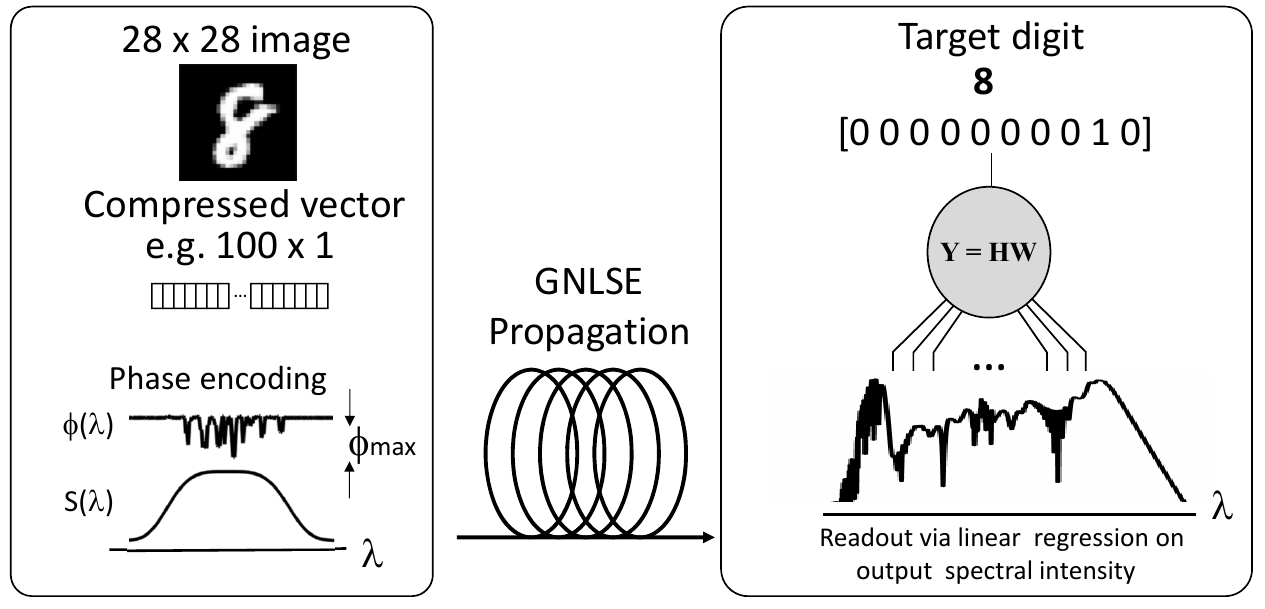}
\caption{Schematic of the supercontinuum ELM model showing the three steps of encoding, generalized nonlinear Schr\"{o}dinger equation (GNLSE) propagation, and spectral readout.}
\label{fig:Fig1}
\end{figure}

We begin by describing the overall system in Fig. 1. An ELM can be considered as a neural network model computing output $\mathbf{Y}$ from input $\mathbf{X}$. An ELM's ability to compute depends on the projection of input $\mathbf{X}$ into a higher-dimensional space via a nonlinear transformation $\mathbf{H} = f_\mathrm{NL}(\mathbf{X}),$ with $\mathbf{H}$ considered as a hidden layer.  ELMs are computationally extremely efficient because it is only the output weights that are trained and, in contrast to deep neural networks or reservoir computers, they do not involve backpropagation or recurrence. Specifically, based on a training dataset $\mathbf{H}$ (computed from $\mathbf{X}$) and corresponding target data $\mathbf{Y}^{\mathrm{T}}$, the ELM learns a model $\mathbf{Y}^{\mathrm{T}} \approx \mathbf{H} \mathbf{W^{\mathrm{out}}}$. Importantly, $\mathbf{W}^\mathrm{out}$ can be computed in a single step according to $\mathbf{W}^\mathrm{out}
\approx \mathbf{H}^{\dagger} \, \mathbf{Y}^\mathrm{T} $,where $\mathbf{H}^{\dagger}$ is the Moore–Penrose pseudoinverse of $\mathbf{H}$. We use pseudoinverse computation here because it solves the linear regression problem directly (minimizing mean squared error between predicted and target outputs) and does not require case-by-case optimization across the wide range of parameters that we study. We discuss the effect of regularization on this regression step further in the Supplemental Document. 

We now consider how this system is applied to handwritten digit classification. For each of the $60,000$ MNIST training images, we perform: (i) bicubic downsampling and flattening from a $28 \times 28$ image to a length-$M$ vector; (ii) encoding of this vector on the spectral phase (or amplitude) of a femtosecond input pulse; (iii) propagation of the encoded pulse in optical fiber; (iv) readout of output spectra into $K$ spectral bins after convolution with a Gaussian spectral response function and addition of a detection noise floor. The downsampling step accounts for the limited spectral bandwidth available for encoding on typical input pulses and, consistent with what would be a necessary approach in any practical system, we assume that spectral measurements are made with a single-shot basis real-time technique \cite{Narhi-2018}. The underlying idea here is that input pulses encoded with different images will produce distinguishable output spectra after nonlinear propagation. Since the spectra for images corresponding to the same digit (0–9) in the training set will be expected to exhibit high-dimensional similarity in their structure, a readout step can be trained to identify these similarities and classify the corresponding digits. 

Specifically, after propagation and readout, the training dataset consists of a $60,000 \times K$ array of spectra, forming the hidden layer $\mathbf{H}$, where the propagation dynamics emulate the ELM transformation $f_\mathrm{NL}$. The target dataset $\mathbf{Y}^\mathrm{T}$ is a $60,000 \times 10$ matrix of known digits (using one-hot encoding), and the $ K \times 10$ weight matrix $\mathbf{W}^\mathrm{out}$ is obtained from solving $\mathbf{W}^\mathrm{out} = \mathbf{H}^{\dagger} \mathbf{Y}^\mathrm{T} $ using the Moore–Penrose
pseudoinverse algorithm.  The model accuracy is determined by applying $\mathbf{W}^\mathrm{out}$ to $\mathbf{H}$ obtained for 10,000 test images not used in training, and computing the accuracy comparing the ELM predictions and the known test digits. Accuracy associated with the training data is also typically calculated. 

The propagation model is the generalized nonlinear Schr\"{o}dinger equation (GNLSE), written in dimensional form as: $ i A_z  - 1/2 \, \beta_2 A_{TT} - i/6 \, \beta_3 A_{TTT} + 1/24 \, \beta_4 A_{TTTT}+ \gamma (1 + i \omega_0 \, \partial_T)(A\,[R \ast |A|^2 ]) = 0 $ \cite{Agrawal-2019}. Here $A(z,T)$ is the complex field envelope (distance $z$, co-moving time $T$), $\beta_k$ are the dispersion coefficients, $\gamma$ is the nonlinearity coefficient, and $\omega_0$ is carrier frequency. The nonlinear response function in the convolution term ($\ast$) is $R(t) = (1-f_R)\delta(t)+f_R h_R(t)$, with Raman fraction $f_R = 0.18$ and $h_R$ the experimental Raman response of fused silica  \cite{Dudley-2006}. Input pulse quantum noise is included via a semiclassical model \cite{Brainis-2005,Dudley-2006} that has been found to yield quantitative agreement with experiment in reproducing supercontinuum noise characteristics \cite{Corwin-2003}. We also included a Raman noise source, but this was found to have negligible influence \cite{Corwin-2003, Dudley-2006}.   For anomalous dispersion regime propagation, we consider dispersion-shifted fiber with 1546.2~nm zero-dispersion wavelength. At pump wavelength 1550~nm, parameters are: $\beta_2 = -0.12 \,\mathrm{ps}^2\mathrm{km}^{-1}$; $\beta_3 = 0.040 \, \mathrm{ps}^3\mathrm{km}^{-1}$; $\beta_4 = 0 \, \mathrm{ps}^4\mathrm{km}^{-1}$; $\gamma = 10.7\,\, \mathrm{W}^{-1}\mathrm{km}^{-1}$. For normal dispersion regime propagation, we consider dispersion-flattened fiber with parameters at 1550~nm: $\beta_2 = 0.987 \,\mathrm{ps}^2\mathrm{km}^{-1}$; $\beta_3 = 7.31 \times 10^{-3}\, \mathrm{ps}^3\mathrm{km}^{-1}$; $\beta_4 = 4.10 \times 10^{-4}\, \mathrm{ps}^4\mathrm{km}^{-1}$; $\gamma = 7.5 \,\, \mathrm{W}^{-1}\mathrm{km}^{-1}$. These parameters correspond to commercially-available fiber. It is straightforward to also include higher-order dispersion, but these truncations yield known characteristic spectral broadening features in both dispersion regimes \cite{Dudley-2006}. At the fiber lengths studied, attenuation at the $\sim$0.5\% level is neglected.  Simulations use a  $2^{11}$ computational grid, but we can analyze the output spectra during readout using different numbers of sampling points. 

Image information is encoded on the input pulses in the frequency domain on a 30~nm full width at half maximum (FWHM) second-order supergaussian spectrum centered on 1550~nm. The FWHM of the corresponding temporal intensity profile is $\Delta \tau \! \sim 182$~fs FWHM. For a particular image, the downsampled image vector of length-$M$ is scaled to a desired phase (or amplitude) modulation depth $\phi_\mathrm{max}$. A 30~nm bandwidth allows encoding with 0.3~nm resolution, consistent with commercial Fourier-domain pulse shaping devices. After encoding, we also scale the input pulse to a particular energy at which we wish to study the dynamics.  

\begin{figure}[hbt]
\centering
\includegraphics[width=0.75\linewidth]{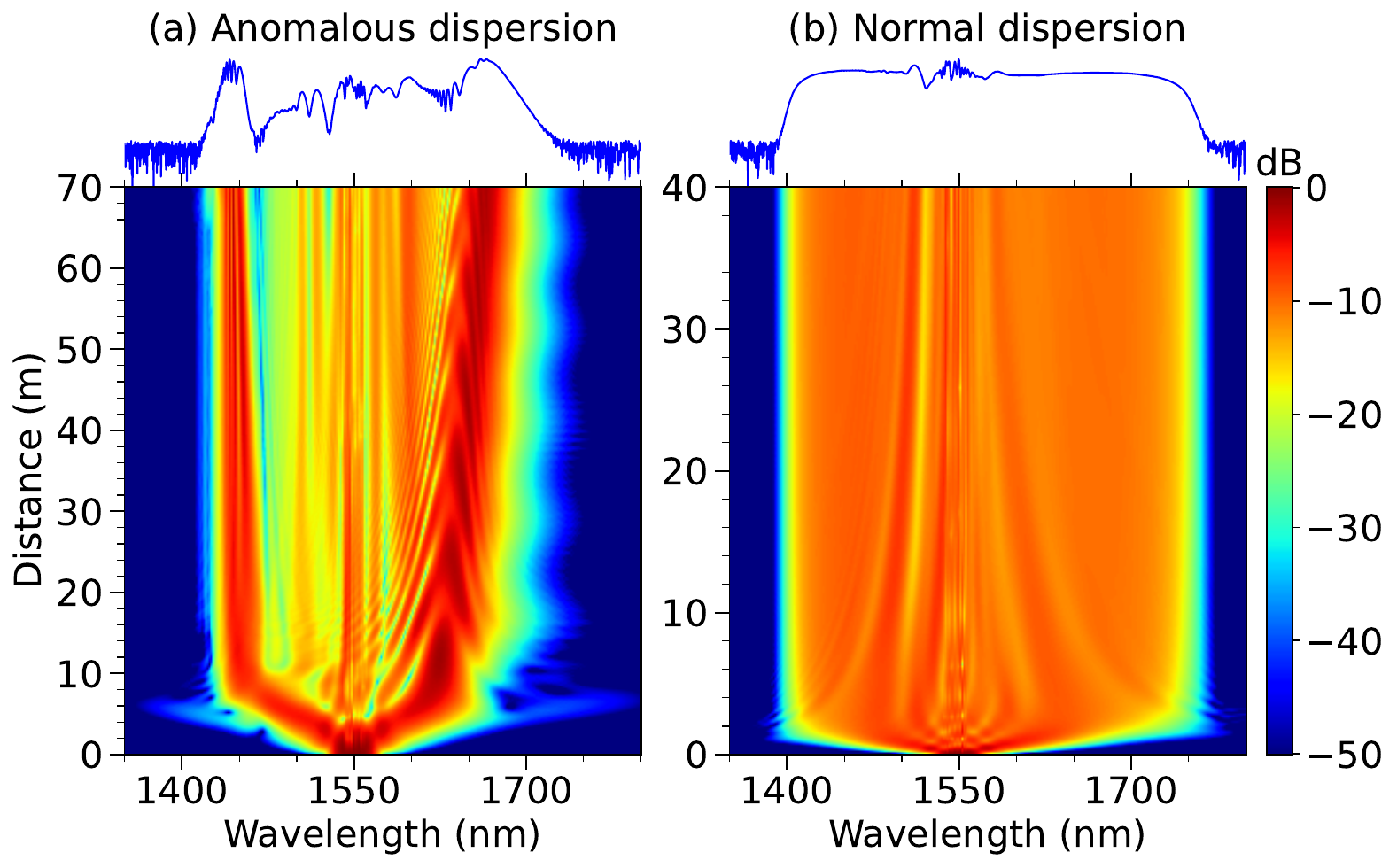}
\caption{Simulated spectral evolution of phase-encoded pulses for (a) anomalous and (b) normal dispersion regime dynamics.} 
\end{figure}

% These results use parameters $\phi_\mathrm{max} = 0.25 \pi$ and $N = 10$.

We first show typical anomalous and normal dispersion regime spectral evolution in Figs. 2(a) and (b). These results use downsampling to $10 \times 10$ ($M = 100$) of one particular image, followed by phase encoding with modulation depth $\phi_\mathrm{max} = 0.25 \pi$. The encoding adds a low-amplitude temporal pedestal at the $\sim$-50~dB level, and for small $\phi_\mathrm{max}$, the corresponding temporal FWHM $\Delta \tau$ is unchanged from that of the unencoded pulse. As a result,  it is convenient to scale the input energy so that the pulse injected in the fiber corresponds to a specified parameter $N = (\gamma P_0 T_0^2 /|\beta_2|)^{1/2}$, where timescale $T_0 \approx \Delta \tau/1.76$. $N$ is a characteristic measure of nonlinear strength which corresponds to soliton number for anomalous dispersion. For anomalous dispersion regime propagation as in Fig.~2(a), input $N = 10$ corresponds to 20.7~pJ energy and $P_0 =103$~W peak power, and we propagate over 70~m. For normal dispersion regime propagation as in Fig.~2(b), $N = 10$ yields $P_0 = 1215$~W peak power and 243~pJ energy, and we propagate over 40~m. To simulate realistic detection, the output spectrum is convolved with a 1~nm Gaussian spectral response (see Supplemental Document) followed by the addition of a -30~dB random noise background to model the finite dynamic range of real-time spectral measurements \cite{Narhi-2018}. The -30~dB noise background was computed relative to the peak of the mean spectral intensity of the ensemble. To generate training and testing spectra under different conditions, we define a particular parameter set and then implement these encoding and propagation steps on each of the 70,000 images.   

Figures 3 and 4 explore parameter dependencies for anomalous and normal dispersion regime propagation respectively. Specifically, Fig.~3 shows results for $10 \times 10$ downsampling, phase-encoding with $\phi_\mathrm{max} = 0.25 \pi$,  and 70~m fiber. Figure 3(a) shows how training (red) and testing (blue) accuracy varies with soliton number $N$ over the range 0.5-13. Figure 3(b) shows results for fixed $N = 10$, but for varying fiber length. Test and training results are compared with benchmark linear regression (green dashed line) based on pseudoinverse computation without fiber propagation. See Supplemental Document for associated confusion charts.  These plots show the expected decrease in accuracy between training and testing, and also the increase in accuracy exceeding 90\%  with increasing propagation complexity (higher $N$ and longer distance.) The decrease in accuracy seen in Figs. 3(a) and 3(b) around $N = 4$ and 10~m respectively, is associated with the fact that we are in the onset phase of soliton fission dynamics (see Supplemental Document).

Figures 3(c) and (d) study aspects of input encoding for $N = 10$ and 70~m fiber. For example, for $10 \times 10$ downsampling, Fig. 3(c) shows how training (red) and testing (blue) accuracy depends on modulation depth $\phi_\mathrm{max}$. There is a clear optimal around $\phi_\mathrm{max} \sim\,0.25 \pi$, and performance away from this point is degraded. Decreasing accuracy for lower modulation depth is expected since image information is weakly encoded and will have limited effect on propagation. The decrease at higher modulation depth arises because greater $\phi_\mathrm{max}$ increasingly modifies the temporal input pulse, reducing peak power at the expense of a low amplitude pedestal. This results in reduced spectral broadening. In Fig. 3(d), we apply optimal modulation $\phi_\mathrm{max} = 0.25 \pi$ and we study how training (red) and testing (blue) accuracies depend on image downsampling i.e. the length $M$ vector describing a $\sqrt{M} \times \sqrt{M}$ image. We see that increasing resolution yields improved results, but test accuracy approaching 90\% can be attained even with only $7 \times 7$ downsampling, representing only $\sim$6\% of the pixels in the original image.  

\begin{figure}[ht!]
\centering
\includegraphics[width=0.75\linewidth]{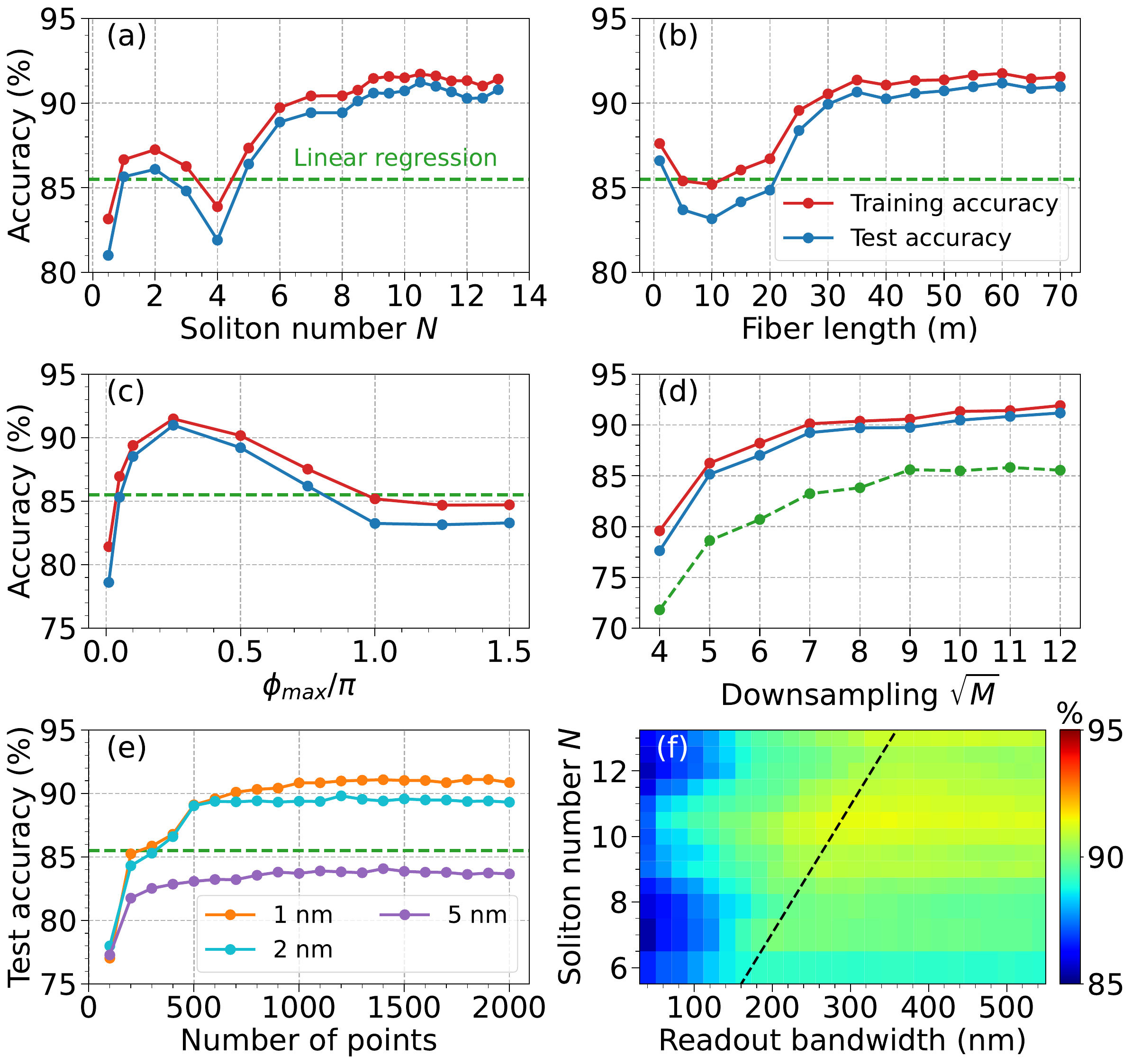}
\caption{Anomalous dispersion results. For $10 \times 10$ downsampling, $\phi_\mathrm{max} = 0.25 \pi$, and 70~m fiber, training (red) and test (blue) accuracy is shown vs: (a) soliton number $N$; (b) fiber length. Green dashed line: linear benchmark. Figures (c-d) use the same color code. For $N = 10$ and 70~m fiber, accuracy is shown: (c) for different $\phi_\mathrm{max}$, and (d) for input downsampling to $\sqrt{M} \times \sqrt{M}$. (e) Test accuracy varying readout bins and convolution bandwidth as indicated. (f) False color plot of test accuracy dependence on $N$ and readout bandwidth (around 1550~nm). Dashed line:  $-20$~dB output  bandwidth.}
\label{fig:3}
\end{figure}

The results in Figs. 3(a)-(d) use readout over the full output spectrum (2048 points over 1317-1882~nm), convolved with a 1~nm spectral response, and with a -30~dB spectral noise floor. It is important to consider how readout parameters influence performance, and these results are shown in Figs. 3(e-f). For downsampling to $10 \times 10$, $\phi_\mathrm{max} = 0.25 \pi$, and $N = 10$, Fig. 3(e) plots test accuracy reading out the spectrum over the full wavelength span, but changing the readout sampling density using different numbers (100-2000) of equispaced bins. We also compare results convolving with three different spectral responses of FWHM: 1~nm (orange), 2~nm (blue), 5~nm (purple). Clearly, a higher sampling density yields improved accuracy, but with 1~nm resolution, $\sim$90\% accuracy can still be approached with only 700 bins. Figure 3(f) examines how the accuracy depends on the value of the readout bandwidth with respect to the overall output spectral bandwidth, using the same $10 \times 10$ downsampling, $\phi_\mathrm{max} = 0.25 \pi$, and 1~nm convolution. The idea here is to study whether we need to read out spectral information over the full bandwidth of the output spectrum, or whether a reduced readout bandwidth is sufficient. The false-color plot shows how test accuracy varies with readout bandwidth (centered on 1550~nm) whilst varying $N$ over 6-13. Of course, as $N$ increases, the bandwidth of the output spectrum will also increase, and this bandwidth (-20~dB level) is shown as the dashed line in the figure.  It is clear that readout over only a portion of the output spectra (i.e. left of the dashed line) leads to reduced accuracy, but once we capture the full output spectral bandwidth i.e. (right of the dashed line) then accuracy improves (then saturates). This is consistent with the interpretation that nonlinear propagation transforms the encoded image information (initially restricted only to 30~nm around the pump) into the high-dimensional space associated with the broadened spectrum.  

\begin{figure}[ht!]
\centering
\includegraphics[width=0.75\linewidth]{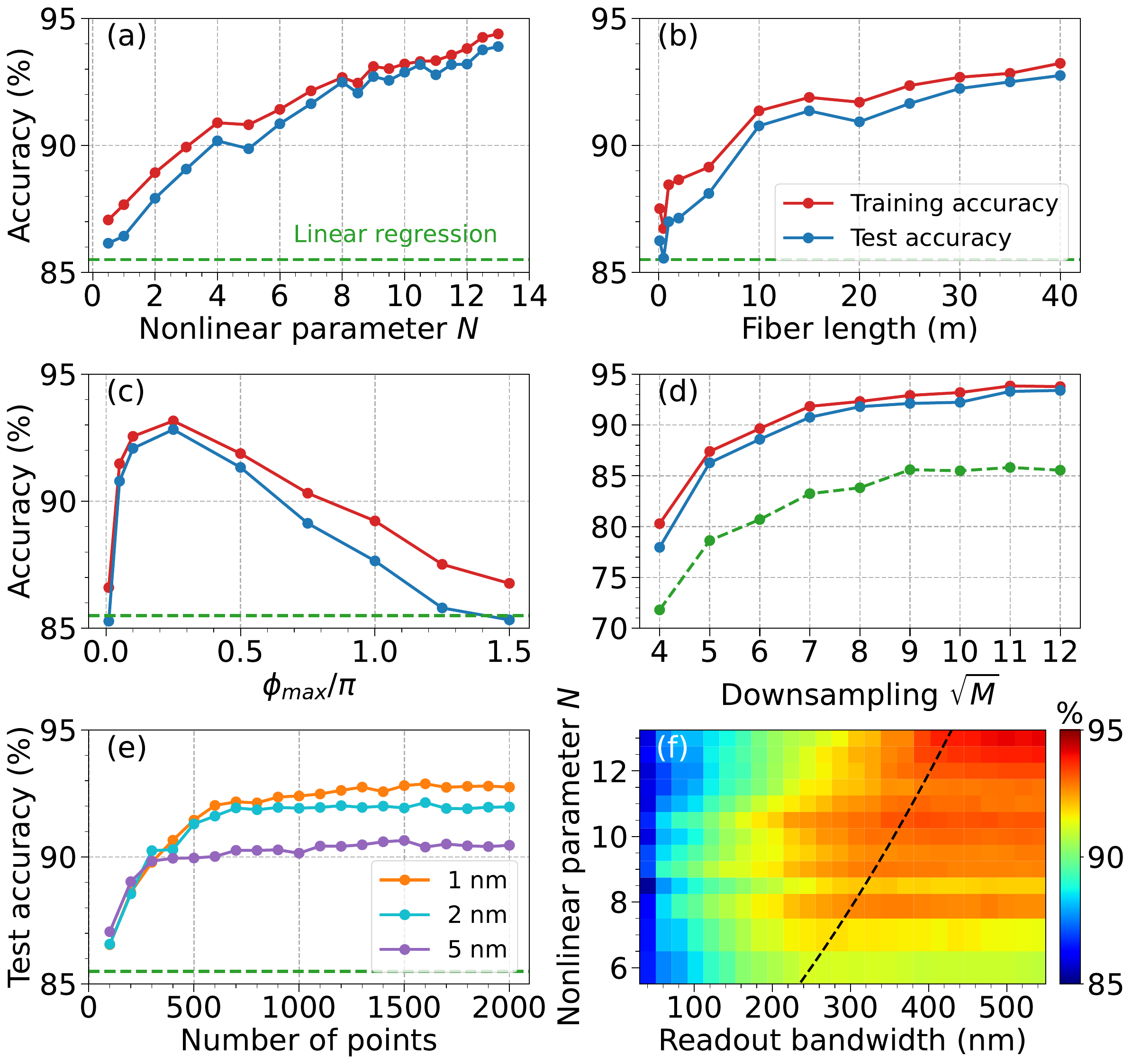}
\caption{Normal dispersion results. For $10 \times 10$ downsampling, $\phi_\mathrm{max} = 0.25 \pi$, and 40~m fiber,  training (red) and test (blue) accuracy is shown vs: (a) nonlinear parameter $N$; (b) fiber length. Green dashed line: linear benchmark. Figures (c-d) use the same color code. For $N = 10$ and 40~m fiber, accuracy is shown: (c) for different $\phi_\mathrm{max}$, and (d) for input downsampling to $\sqrt{M} \times \sqrt{M}$. (e) Test accuracy varying readout bins and convolution bandwidth as indicated. (f) False color plot of test accuracy dependence on $N$ and readout bandwidth (around 1550~nm). Dashed line:  $-20$~dB output bandwidth.}
\label{fig:4}
\end{figure}

Figures 4(a)-(f) show  results for normal dispersion regime propagation. The overall trends and dependence on parameters such as $N$ and fiber length are qualitatively alike, although normal dispersion regime propagation consistently shows $\sim$3\% improvement in training and testing accuracy. This can be attributed to the well-known observation of improved noise characteristics in the normal dispersion regime \cite{Agrawal-2019}. However, for both anomalous and normal dispersion, we anticipate that accuracy would ultimately degrade for increased power and/or fiber length as a result of deleterious effects such as incoherent supercontinuum dynamics, polarization instabilities etc \cite{Agrawal-2019}. 

The studies in Figs. 3 and 4 were repeated using spectral amplitude encoding on the input pulses. This involved multiplying the supergaussian input spectrum by an amplitude encoding mask with variable modulation depth. The general trends and results (see Supplemental Document) were extremely similar to those obtained with phase encoding in terms of parameter dependence and classification accuracy. 

In Fig. 5 we study the impact of noise on ELM performance using phase encoding for $\phi_\mathrm{max} = 0.25 \pi$ and $N=10$. Firstly, we recall that the results in Figs. 2-4 were obtained with quantum noise on the input pulses and a -30~dB readout noise background. It is straightforward with simulations to model the ELM without input quantum noise or a readout noise floor, and this allows us to study the ideal mathematical properties of the GNLSE to act as a nonlinear ELM transfer function.  These results are shown as the crosses in Fig.~5 and indicate what can be considered upper limit ideal test accuracies of 96.7\% and 95.0\% for normal and anomalous dispersion regime propagation respectively. 
 Significantly, the addition of only input quantum noise (i.e. again with no imposed readout noise floor) reduces these ideal test accuracies to 94.8\% and 91.9\% respectively, an important result that highlights an intrinsic quantum noise penalty for this class of fiber-based ELM. Of course the imposed -30~dB instrumental noise floor on the output spectra reduces accuracy further, with the corresponding results in this case $\sim $93\% and $\sim$91\% respectively for anomalous and normal dispersion regime propagation. 

\begin{figure}[hb]
\centering
\includegraphics[width=0.75\linewidth]{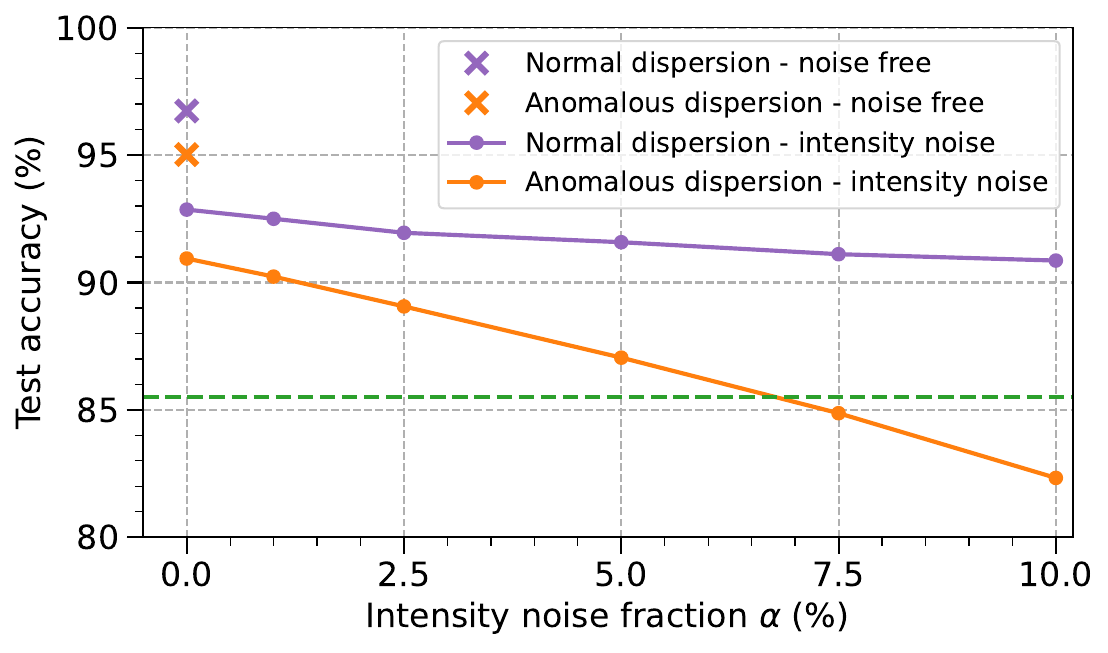}
\caption{Lines: Dependence of ELM test accuracy on multiplicative noise for normal (purple) and anomalous (orange) dispersion regime propagation. These results include input quantum noise and a -30~dB readout noise floor. Crosses: ideal accuracies without any input or readout noise. Green dashed line: linear regression benchmark.}
\label{fig:Fig5}
\end{figure}

As might be expected, any additional  input noise degrades performance. To show this, we apply a simple multiplicative intensity noise model of the form: $A_\mathrm{in}(T) = A_\mathrm{0}(T) [1 + \alpha/2 \, \eta(T)] $, where $A_\mathrm{0}$ is the temporal input field after addition of quantum noise and before spectral encoding. Parameter $\alpha$ corresponds to an intensity noise fraction, and $\eta(T)$ is a random variable. Figure 5 shows results applying this noise on input pulses with $10 \times 10$ downsampling, phase encoding, $\phi_\mathrm{max} = 0.25 \pi$ and $ N = 10$.  It is clear that testing accuracy decreases with increasing noise fraction in both anomalous (orange) and normal (purple) dispersion regimes. Results are clearly more degraded in the anomalous dispersion regime, further highlighting the noise sensitivity of anomalous dispersion regime propagation \cite{Agrawal-2019}.

The GNLSE model framework is a powerful and flexible tool to explore nonlinear fiber propagation as applied to ELM systems, and provides important insights for experimental design. Consistent with recent experiments, our results suggest that normal dispersion propagation yields improved accuracy and reduced noise sensitivity, but satisfactory results can still be obtained with anomalous dispersion regime propagation for suitable parameters \cite{Fischer-2023,Lee-2024,Hary-2025,Saeed-2025,Muda-2025}. A further key conclusion is to identify that input pulse quantum noise will likely impose an intrinsic penalty for all nonlinear fiber propagation-based ELMs.

\section*{Funding} Agence Nationale de la Recherche (ANR-15-IDEX-0003, ANR-17-EURE-0002, ANR-20-CE30-0004); Institut Universitaire de France; ERC Consolidator grant INSPIRE (101044777); Research Council of Finland (368650).

\section*{Disclosures} The authors declare no conflicts of interest.

\section*{Data availability} Data underlying the results presented in this paper may be obtained from the authors upon reasonable request.

\newpage

\section*{Supplemental Document}

\section{Additional information on methods}
This document presents additional material relating to the performance of an extreme learning machine (ELM) based on nonlinear and dispersive optical fiber propagation. This includes: (i) Further information and discussion about methods; (ii) Results showing confusion charts for anomalous and normal dispersion regime propagation applied to the MNIST digits dataset; (iii) Analysis of the MNIST Fashion dataset using phase-encoding; (iv) Analysis of the MNIST Digit dataset using amplitude encoding;  (v) Additional discussion and consideration of future research directions. 

\subsection{Simulation Model}

As discussed in the main manuscript, the propagation model used is the generalized nonlinear Schr\"{o}dinger equation (GNLSE), written in dimensional form \cite{Agrawal-2019} as:
\begin{equation}
i \frac{\partial A}{\partial z} - \frac{1}{2} \, \beta_2 \frac{\partial^2 A}{\partial T^2} - \frac{i}{6} \, \beta_3 \frac{\partial^3 A}{\partial T^3} + \frac{1}{24} \, \beta_4\frac{\partial ^4 A}{\partial T^4}+ \gamma \left( 1 + i \omega_0 \, \frac{\partial}{\partial T} \right) \left( A\,[R \ast |A|^2 ]\right) = 0. 
\end{equation}
Here $A(z,T)$ is the complex field envelope as a function of distance $z$ and co-moving time $T$), $\beta_k$ are the dispersion coefficients, $\gamma$ is the nonlinearity coefficient, and $\omega_0$ is carrier frequency. The nonlinear response function in the convolution term ($\ast$) is $R(t) = (1-f_R)\delta(t)+f_R h_R(t)$, with Raman fraction $f_R = 0.18$ and $h_R$ the experimental Raman response of fused silica  \cite{Dudley-2006}. Input pulse quantum noise is included via a semiclassical model \cite{Brainis-2005,Dudley-2006} that has been previously found to yield quantitative agreement with experiment in reproducing supercontinuum noise characteristics \cite{Corwin-2003}. We also included a Langevin Raman noise source, but this was found to have negligible influence, consistent with previous studies of intensity noise in supercontinuum generation \cite{Corwin-2003}. This GNLSE model is a standard model of the field and has been successfully used in many studies of fiber propagation in both anomalous and normal dispersion regimes.  

A limiting factor in any numerical simulation is the need to avoid numerical artifacts, and for GNLSE modeling, a key requirement is to ensure that temporal and spectral structure remains confined to the corresponding computational window.  In our simulations using a  $2^{11}$-length grid, this limits us to study the propagation of pulses up to nonlinear parameter $N = 13$.  This restriction can be relaxed using more temporal-spectral grid points, as well as more integration points in the $z$-dimension, but with the drawback of significantly increased computation time. In a machine learning training context, this is a major constraint giving the large training sets and the need to explore performance over a broad range of parameters.  As a specific example, simulations modeling propagation with $N = 15$ require that we increase the number of grid points by a factor of 4, and increase the number of integration points by a factor of 2, which is a factor of 8 increase in overall computation time. From a physical perspective, however, we note that we would expect some degree of saturation and ultimately a decrease in accuracy with increasing $N$ as we see the onset of effects such as modulation instability and decoherence in the anomalous dispersion regime and polarization instabilities in the normal dispersion regime.  

\subsection{Choice of Fiber Types and Parameters}

Our choice of simulation parameters is motivated by two factors: (i) we wish to choose experimentally realistic parameters based on commercially available sources and optical fiber; (ii) we wish to contrast ELM performance for the two main physical regimes of fiber propagation: supercontinuum generation with pumping in the anomalous dispersion regime close to the fiber zero dispersion wavelength, and self-phase modulation (SPM)-optical wavebreaking dynamics with a pump in the normal dispersion regime far from the fiber zero dispersion wavelength (ZDW) \cite{Agrawal-2019}. The physical processes underlying these spectral broadening mechanisms are distinct and lead to very different output spectral structure: supercontinuum generation typically sees a complex spectrum with multiple peaks and significant spectral content on either side of the ZDW, whereas SPM and optical wavebreaking dynamics yield a more uniform spectrum that lies entirely in the normal dispersion regime. 

The fiber dispersion curves associated with these two regimes of spectral broadening are very different. Fibers used in supercontinuum generation typically have a monotonically-decreasing variation of $\beta_2$ with respect to wavelength that changes sign across the ZDW. In our case we choose a fiber with ZDW of 1546.2~nm, specifically a dispersion-shifted fiber with parameters at 1550~nm of: $\beta_2 = -0.12 \,\mathrm{ps}^2\mathrm{km}^{-1}$; $\beta_3 = 0.040 \, \mathrm{ps}^3\mathrm{km}^{-1}$; $\beta_4 = 0 \, \mathrm{ps}^4\mathrm{km}^{-1}$; $\gamma = 10.7\,\, \mathrm{W}^{-1}\mathrm{km}^{-1}$. In contrast, the fiber choice for SPM/optical wavebreaking should present all normal dispersion across the target spectrum, and we select a C-band dispersion-flattened fiber where $\beta_2$ has a concave-up profile with wavelength, and where $\beta_2 > 0 $ across the full broadened spectral bandwidth.  In this case, parameters at 1550~nm were: $\beta_2 = 0.987 \,\mathrm{ps}^2\mathrm{km}^{-1}$; $\beta_3 = 7.31 \times 10^{-3}\, \mathrm{ps}^3\mathrm{km}^{-1}$; $\beta_4 = 4.10 \times 10^{-4}\, \mathrm{ps}^4\mathrm{km}^{-1}$; $\gamma = 7.5 \,\, \mathrm{W}^{-1}\mathrm{km}^{-1}$.   

With our selection of fiber and source parameters, the generated spectra have comparable bandwidths, but it is important to stress that the underlying evolution dynamics are very different in the two cases.  In this context, it is common to introduce characteristic nonlinear and dispersive  length scales $L_\mathrm{NL} = (\gamma P_0)^{-1}$ and $L_D = T_0^2/|\beta_2|$ respectively, where $P_0$ and $T_0$ are the parameters of an input pulse $A(0,T) = \sqrt{P_0}\,\mathrm{sech}(T/T_0)$. Dimensionless $N = (L_\mathrm{D}/L_\mathrm{NL})^{0.5} = (\gamma P_0 T_0^2 / |\beta_2|)^{0.5} $ has the interpretation as the soliton number for anomalous dispersion, and in the normal dispersion regime as a nonlinear parameter combining fiber and source parameters. Indeed in Figs. 3(a) and 4(a) of the main manuscript, we used $N$ as a normalized parameter to compare anomalous and normal dispersion regime performance for increasing pulse energy.  The characteristic nonlinear and dispersive interaction length $L_\mathrm{c} = \sqrt{L_\mathrm{NL}\,L_\mathrm{D}} $ is a convenient measure to describe the onset of key dynamical processes in both the anomalous \cite{Dudley-2006} and normal \cite{Finot-2008} dispersion regimes.  

It is useful to discuss the dynamics of spectral broadening in the anomalous and normal dispersion regimes in terms of the characteristic length scales for the specific case $N = 10$. This corresponds to the evolution plots in Fig.~2 of the main manuscript. In the case of anomalous dispersion regime pumping and the supercontinuum dynamics shown in Fig.~2(a), the characteristic interaction length $L_\mathrm{c} \approx 8.6$~m is associated with the onset of soliton fission. In this case (see Fig. 3(b)), we find that a fiber length of 70~m (around $8 L_\mathrm{c}$) yields high accuracy, physically associated with a sufficient distance to see the expected dynamics of dispersive wave generation and the Raman soliton self-frequency shift \cite{Dudley-2006,Agrawal-2019}.  In the case of normal dispersion regime pumping and the SPM-wavebreaking dynamics in Fig.~2(b), the characteristic interaction length $L_\mathrm{c} \approx 1 $~m describes the onset of optical wavebreaking \cite{Anderson-1992,Finot-2008}. Here (see Fig. 4(b)), we enter into a regime of high accuracy at a distance of 10~m (around $10 L_\mathrm{c}$). This corresponds to the development of significant uniform spectral structure on either side of the pump \cite{Finot-2008,Agrawal-2019}, and a comparable overall bandwidth to the anomalous dispersion regime case. We simulate propagation in the normal dispersion regime out to a total length of 40~m to study the potential for improved accuracy for greater fiber lengths.  Note, however, that similar to the discussion above in Section 1B relating to the effect of increasing $N$, we would also expect some degree of saturation and ultimately a decrease in accuracy with increasing fiber length as we see the onset of effects such as modulation instability and decoherence in the anomalous dispersion regime and polarization instabilities in the normal dispersion regime.

\subsection{Readout methodology}

Figures 3(e) and 4(e) of the main manuscript present results showing how the bandwidth of the spectral response function used at the readout step impacts the test accuracy of the ELM.  We now provide additional description of this response function. We first note that the functional form of the response will depend on the particular experimental method used to record the spectra, resulting from the convolution of the system dispersive and detection responses. Each experimental setup will be different – for example, a scanning aperture based spectrometer would be expected to yield a $\mathrm{sinc}^2$ or Airy function response with side lobes (depending on the nature of the aperture used), whereas a propagation-based dispersive Fourier transform \cite{Goda-2013,Godin-2022} setup would be expected to have a response closer to a Gaussian, possibly also with asymmetric ringing depending on the photodetector used and the oscilloscope bandwidth.  With suitable design and choice of setup, low amplitude features of the spectral response can generally be neglected, but what cannot be neglected is the spectral resolution. Mathematically, the spectral resolution will determine the number of independent regions of the spectrum that can be used for readout and thus the effective higher-order dimensionality of system. In our simulations, for generality and reproducibility when studying the effect of spectral resolution, we selected a Gaussian function as the response for the analysis presented in Figs 3(e) and 4(e). The values of resolution bandwidths were taken as corresponding to typical spectral measurement techniques used around 1550~nm.  

Concerning bin selection during readout, we used a deterministic binning strategy dividing the output spectrum into uniformly spaced intervals without randomization. This was to ensure an identical and reproducible strategy as we studied the dependence of accuracy on the different physical parameters. In Figs 3(e) and 4(e) of the main manuscript when we vary the number of readout bins from 100-2000, these bins are distributed uniformly across the spectral window of 1317-1882 nm, with the bin spacing varying from 5.6~nm for 100 bins to 0.28~nm for 2000 bins.  In Figs 3(f) and 4(f) of the main manuscript when we study the effect of readout bandwidth around the pump wavelength of 1550~nm, we use a fixed bin separation of 0.28~nm and as the readout span increases from 40-540~nm, this correspondingly increases the number of bins from 150 to 1975. 

An important aspect of the readout step is regression of the hidden layer matrix $\mathbf{H}$. In the results shown in the main manuscript, we used the Moore-Penrose pseudoinverse which directly solves the regression problem by minimizing the mean squared error between predicted and target outputs.  This method was chosen because we wished to study ELM performance over a wide range of physical parameters, independent of tunable hyperparameters optimized during the offline readout. There are of course other regression techniques available, and the use of ridge regression in particular has potential to improve generalization by tuning a regularization parameter $\lambda_\mathrm{R}$ through:  $\mathbf{W}^{\text{out}} = (\mathbf{H}^\top \mathbf{H} + \lambda_\mathrm{R} I)^{-1} \mathbf{H}^\top \mathbf{Y}^T$. (Note that $\top$ indicates transpose and the superscript $T$ on $\mathbf{Y}$ does not refer to a transpose operation but rather indicates that this is the training data.)  However, we found that ridge regression yielded little improvement for our parameter sets.  In fact, this might be anticipated in our case since the datasets we use (60,000) are large compared to the number of features/hidden layer neurons (maximum of 2048).  For completeness, Fig.~S1 plots the test accuracy obtained using pseudoinverse and ridge regression (scanning the regularization parameter) for the MNIST digit dataset for both normal (40~m fiber) and anomalous (70~m fiber) dispersion as indicated in the caption. Ridge regression slightly outperforms pseudoinverse regression for some values of $\lambda_\mathrm{R}$, but this small level of improvement ($< 0.3\%$ in the anomalous dispersion regime, $< 0.1\%$ in the normal dispersion regime) suggests that the pseudoinverse solution is close to optimal for generalization. We found similar results and trends for the MNIST Fashion dataset.  

\begin{figure*}[h]
\centering
\includegraphics[width=0.85\linewidth]{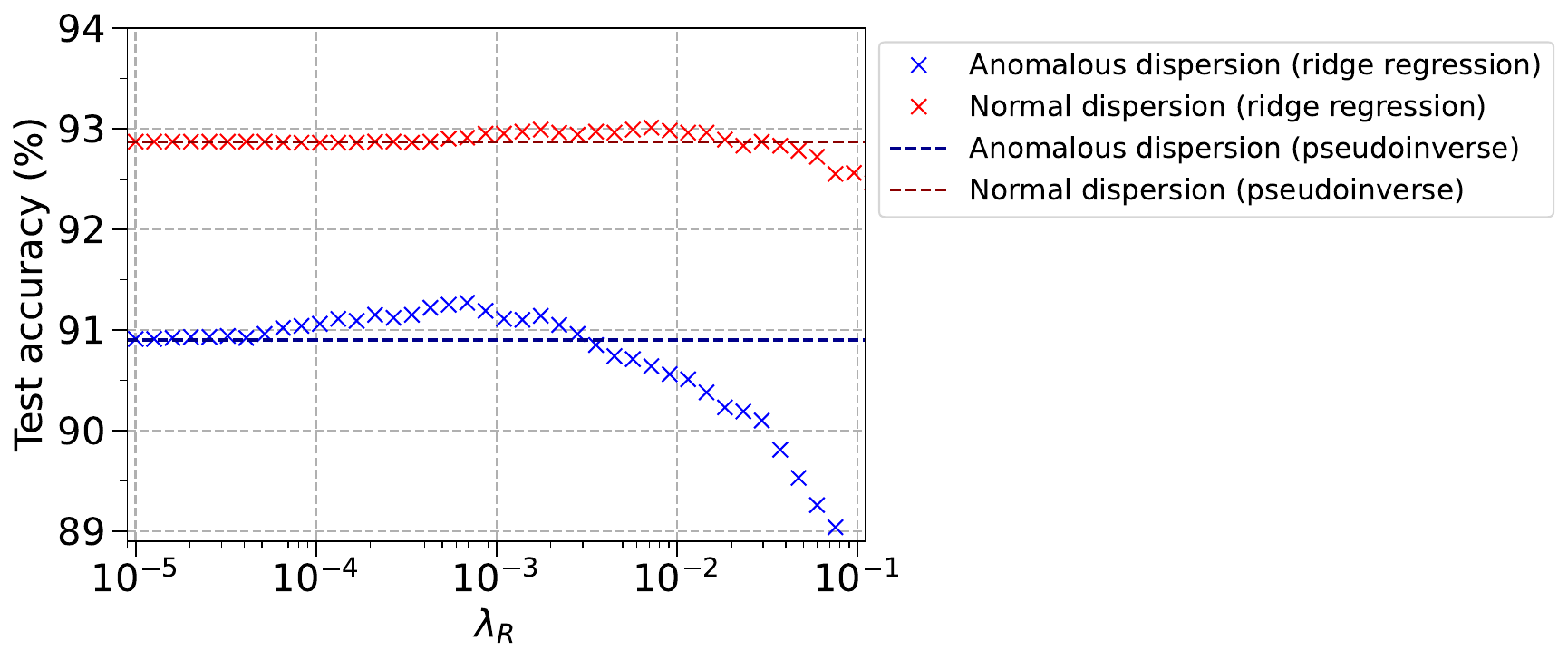}
\caption{Results for the MNIST Digit dataset, comparing test accuracy when using ridge regression and the pseudoinverse.  We used $10 \times 10$ downsampling and $N = 10$ and the figure plots results for normal (red) and anomalous (blue) dispersion regime propagation. Ridge regression results are shown as crosses and the baseline pseudoinverse result is shown as the dashed line in each case.}
\label{fig:ridge}
\end{figure*}

\section{Additional results and discussion}

\subsection{Accuracy dependence on soliton number and distance for supercontinuum propagation}

For the case of supercontinuum propagation with an anomalous dispersion regime pump, the results in Fig 3(a) of the main manuscript show a decrease in accuracy for MNIST digit classification around soliton number $N = 4$, and the results in Fig 3(b) show a decrease in accuracy around 10~m propagation distance.  To interpret this, we note that with low soliton numbers or shorter propagation distances, we are in the emergent phase of the soliton fission process \cite{Agrawal-2019} such that, from the perspective of an ELM, the nonlinear dynamics have not yet sufficiently projected the input information into the high-dimensional spectral space associated with the supercontinuum, and so readout over the full bandwidth will include large spectral regions that contribute little or not at all to the computational transformation and dimensionality expansion of the input images.  

\subsection{MNIST Digit dataset - Phase Encoding Confusion Charts}

To complement the results in Figs 3 and 4 of the main manuscript, it is useful to show confusion matrices (charts) associated with the output layer readout.  In particular, for $10 \times 10$ downsampling of the MNIST digit dataset and phase encoding with $\phi_\mathrm{max} = 0.25$, Fig. S2 shows the confusion charts corresponding to: (a) benchmark linear regression, (b) anomalous dispersion regime propagation for $N = 10$ and 70~m fiber, (c) normal dispersion regime propagation for $N = 10$ and 40~m fiber. The choice of $N = 10$ here is because this corresponds to a best-case result for anomalous dispersion, and we keep this constant for the comparison with the normal dispersion regime results.  

\begin{figure*}[ht]
\centering
\includegraphics[width=0.8\linewidth]{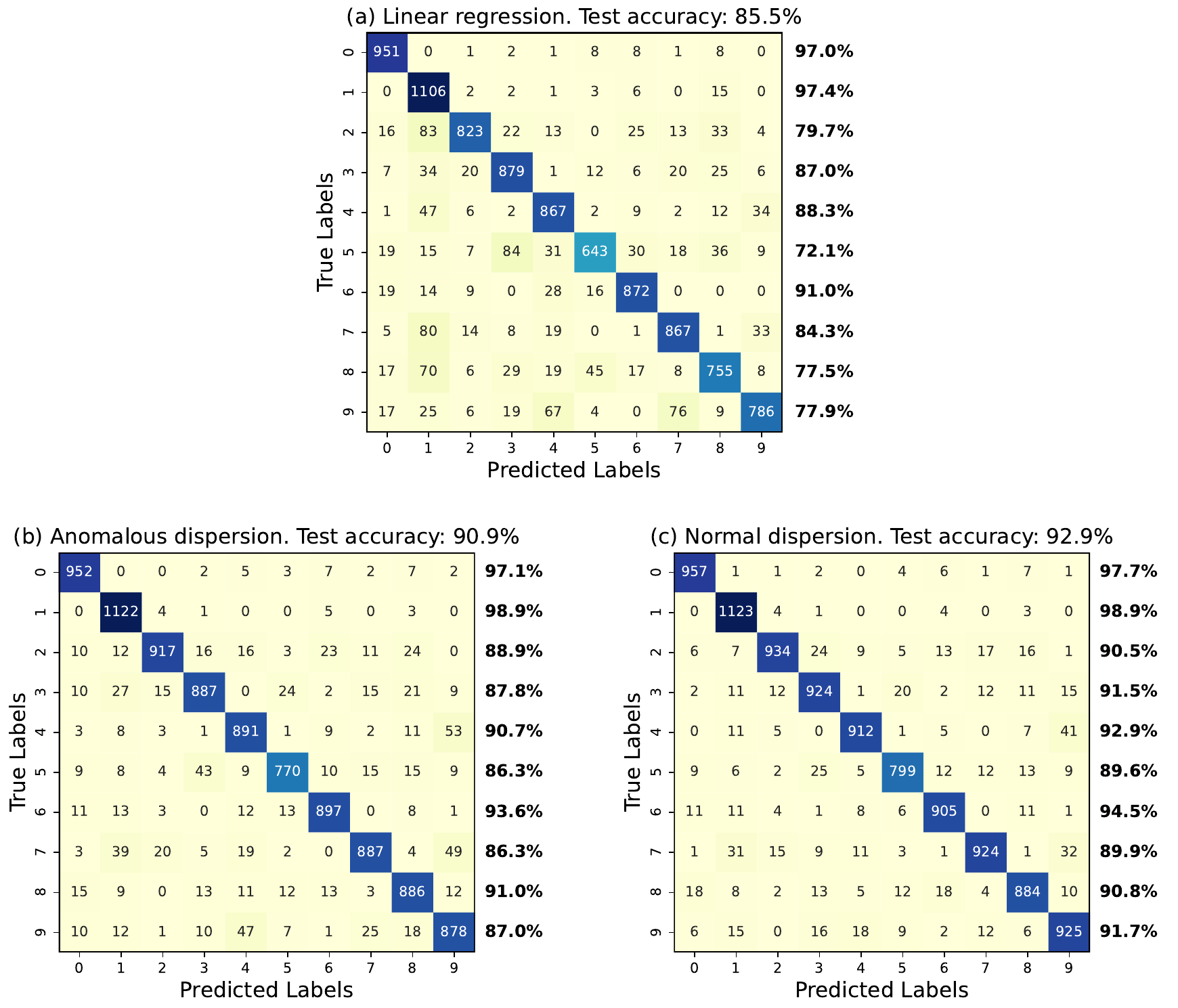}
\caption{For $10 \times 10$ downsampling of the MNIST digit dataset and phase encoding with $\phi_\mathrm{max} = 0.25$, the figure shows confusion charts for: (a) benchmark linear regression, (b) anomalous dispersion regime propagation for $N = 10$ and 70~m fiber, (c) normal dispersion regime propagation for $N = 10$ and 40~m fiber.  The numbers to the right of each row is the digit-by-digit accuracy, and the title of each chart gives the accuracy computed across the full dataset.}
\label{fig:confusion}
\end{figure*}

\subsection{MNIST Digit results - Amplitude Encoding}
\begin{figure*}[ht!]
\centering
\includegraphics[width=0.75\linewidth]{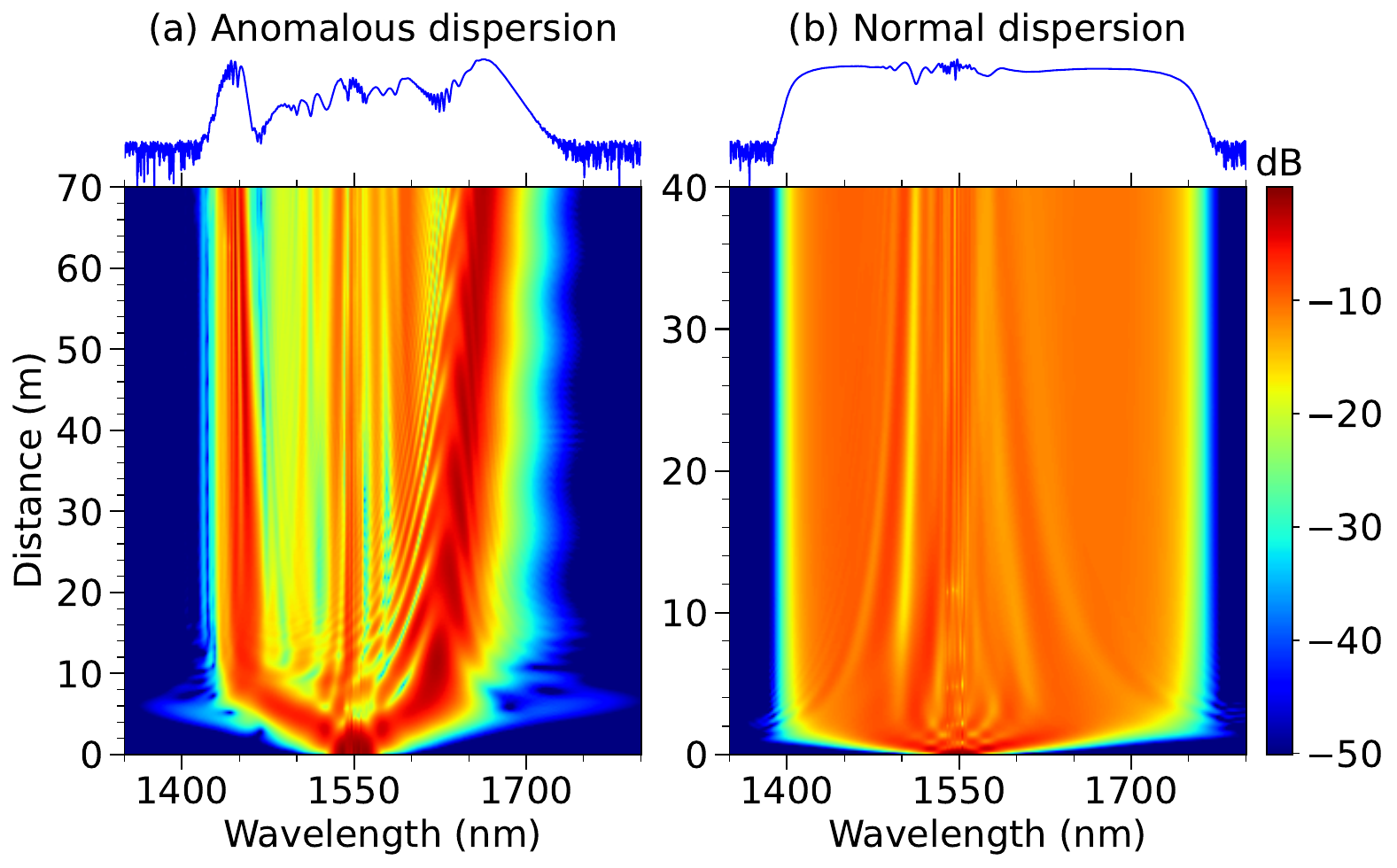}
\caption{Spectral evolution with amplitude-encoding: (a) anomalous and (b) normal dispersion.} 
\label{fig:1s}
\end{figure*}

\begin{figure*}[ht!]
\centering
\includegraphics[width=0.75\linewidth]{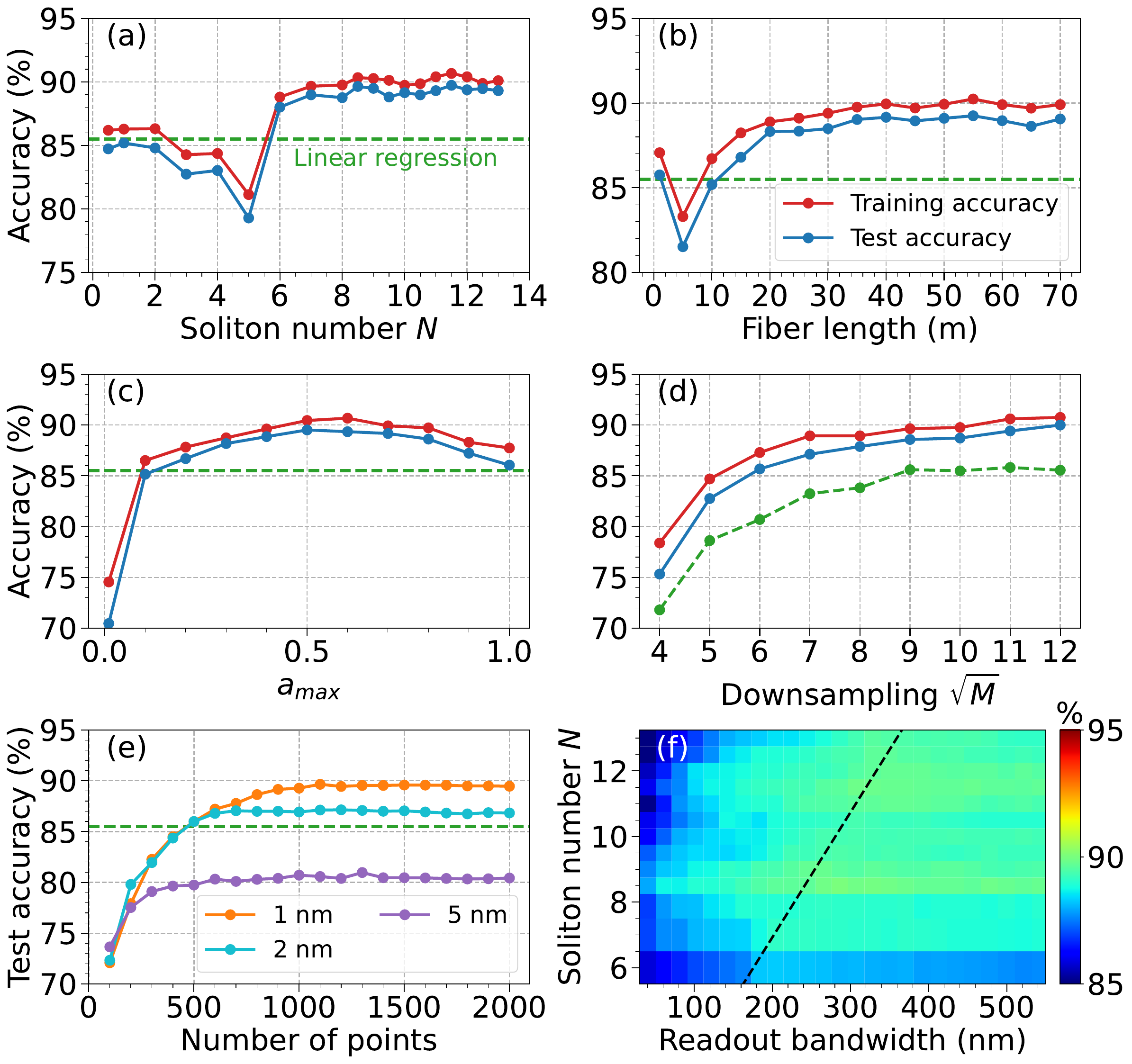}
\caption{MNIST Digit results: amplitude encoding and anomalous dispersion. For $10 \times 10$ downsampling, $a_\mathrm{max} = 0.5 $, and 70~m fiber, training (red) and test (blue) accuracy is shown as a function of (a) soliton number $N$ and (b) fiber length. Green dashed line: linear benchmark. Figures (c-d) use the same color code. For $N = 10$ and 70~m fiber, accuracy is shown: (c) for different $a_\mathrm{max}$, and (d) for downsampling to $\sqrt{M} \times \sqrt{M}$. (e) Test accuracy varying readout bins and convolution bandwidth as indicated. (f) False color plot of test accuracy dependence on $N$ and readout bandwidth. Dashed line:  $-20$~dB output  bandwidth.} 
\label{fig:2s}
\end{figure*}

\begin{figure*}[ht!]
\centering
\includegraphics[width=0.75\linewidth]{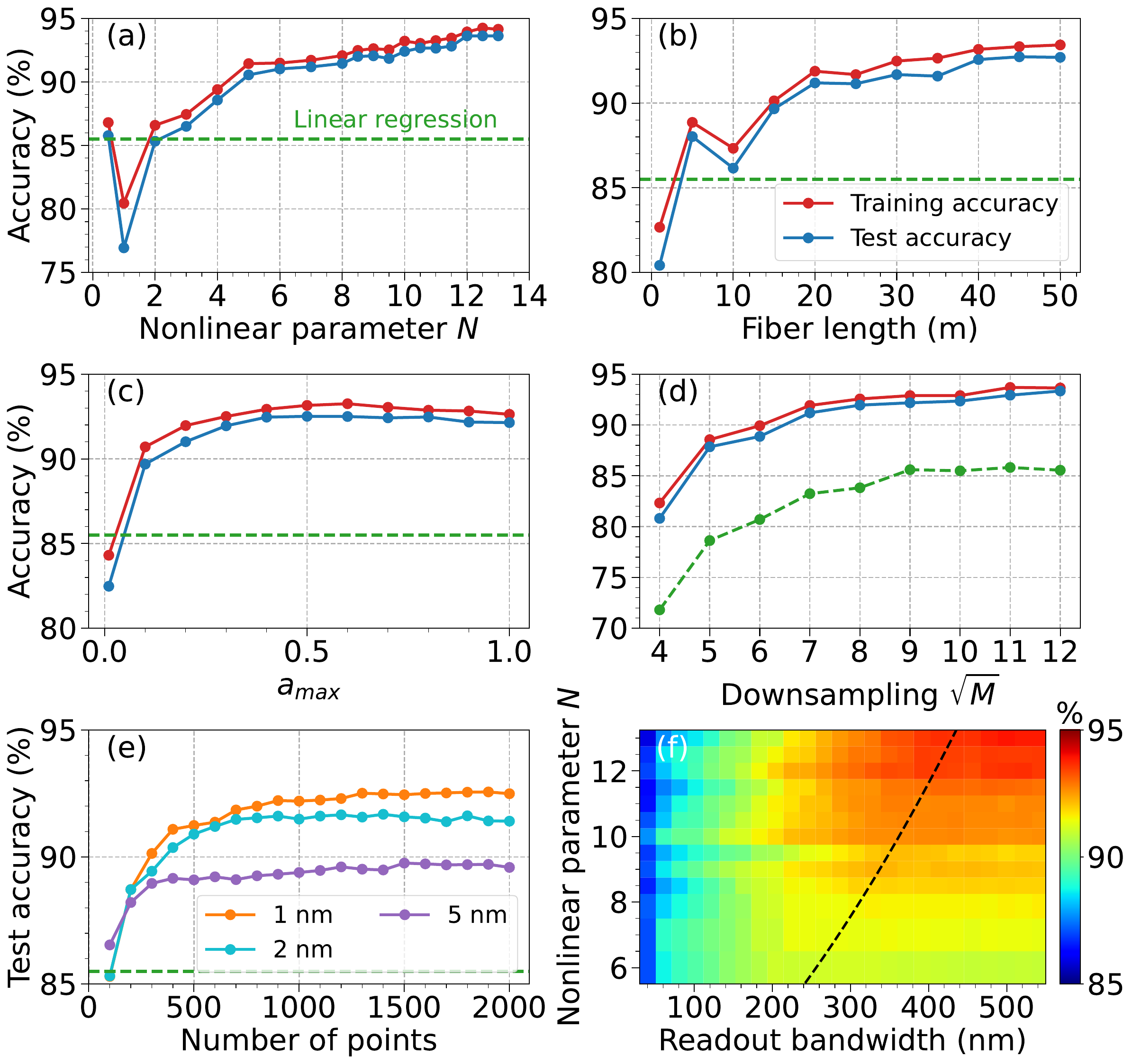}
\caption{MNIST digit results: amplitude encoding and normal dispersion. For $10 \times 10$ downsampling, $a_\mathrm{max} = 0.5$, and 40~m fiber,  training (red) and test (blue) accuracy is shown as a function of (a) nonlinear parameter $N$ and (b) fiber length. Green dashed line: linear benchmark. Figures (c-d) use the same color code. For $N = 10$ and 70~m fiber, accuracy is shown: (c) for different $a_\mathrm{max}$, and (d) for input downsampling to $\sqrt{M} \times \sqrt{M}$. (e) Test accuracy varying readout bins and convolution bandwidth as indicated. (f) False color plot of test accuracy dependence on $N$ and readout bandwidth. Dashed line:  $-20$~dB output bandwidth.} 
\label{fig:3s}
\end{figure*}

Figures~\ref{fig:1s}-\ref{fig:3s} show results similar to those in the main manuscript, but for amplitude encoding applied to the MNIST dataset. Figure S3 shows typical spectral evolution for (a) anomalous and (b) normal dispersion regimes. As in Fig.~2 of the main manuscript, these results use $10 \times 10$ ($M = 100$) downsampling, but now followed by amplitude encoding with modulation depth of $a_\mathrm{max} = 0.5$. For anomalous dispersion regime propagation, input $N = 10$ corresponds to energy of 20.7~pJ and peak power $P_0 = 102$~W, and we propagate over 70~m. For normal dispersion regime propagation, $N = 10$ gives peak power $P_0 =1203$~W and energy of 243~pJ, and we propagate to 40~m. The upper subplots show corresponding output spectra after the convolution with 1~nm Gaussian response and addition of -30~dB noise background. 

Figure~\ref{fig:2s} shows ELM performance for anomalous dispersion regime propagation, with results similar to Fig. 3 of the main manuscript, but this time using amplitude encoding. Figure~\ref{fig:2s} (a) shows dependence of training and test accuracies on input soliton number $N$ compared with the linear regression benchmark (green dashed curve). The general trend remains similar to the phase encoding case [see Fig. 3(a) of the main manuscript], highlighting that to outperform the linear regression benchmark a relatively high nonlinearity ($N > 5$) is required. Meanwhile, in the anomalous dispersion regime, the training and test accuracies are slightly lower for amplitude coding compared to phase encoding. Similar trends can be seen in Fig.~\ref{fig:2s} (b), which plots training and test accuracies as a function of the fiber length. Figure~\ref{fig:2s} (c) shows how the training and test accuracies depend on the amplitude modulation depth $a_\mathrm{max}$, which we vary over the range $0.01$ to $1$. The optimal amplitude encoding depth was found to be $a_\mathrm{max} \sim 0.5$. It is noticeable that the performance of the ELM is less sensitive to the choice of amplitude modulation depth than the phase modulation depth. 

Figure~\ref{fig:2s} (d) shows how the training and test accuracies depend on the downsampling of the MNIST images. Once again, we can see that while the general trend remains similar to that for phase encoding, in the case of amplitude encoding we get around $1\%$ less training and test accuracies. Figure~\ref{fig:2s} (e) shows the test accuracy when reading the spectrum over the full wavelength range, but changing the sampling density of the readout using different numbers of equispaced bins (100-2000). These results are shown for the output spectra convolved with three different spectral responses of FWHM: 1~nm (orange), 2~nm (blue), 5~nm (purple). Finally, Fig.~\ref{fig:2s} (f) shows the test accuracy as a function of the input soliton number $6 \leq N \leq 13$ and the readout bandwidth which we vary from 40 to 540 nm. We note that the results obtained in the case of amplitude coding also suggest that a larger bandwidth should be selected in order to obtain better generalization. This suggests that nonlinear propagation governed by the generalized non-linear Schr\"odinger equation efficiently transforms the encoded image information into the high-dimensional space associated with the extended spectrum in both the phase and amplitude encoding cases.

In Figure~\ref{fig:3s} we present the same analysis as before for the case of normal dispersion regime propagation and amplitude encoding. As in the case of phase encoding (main manuscript) the accuracies obtained with amplitude encoding are higher for propagation in the normal dispersion regime propagation when compared to the anomalous dispersion regime.

\newpage

\section{MNIST Fashion dataset - Phase Encoding}

In addition to results on the MNIST handwritten digit dataset, we also studied the performance of the fiber-based ELM to classify the more complex image data in the MNIST Fashion dataset. Fashion-MNIST provides a more challenging test of machine learning models because of increased visual complexity and feature overlap. The encoding and readout processes were the same as described in the main manuscript for the MNIST digit dataset.  The results in Fig. S6 correspond to the case of anomalous dispersion regime propagation and are to be compared with the results in Fig. 3 in the main manuscript.  The results in Fig. S7  correspond to the the case of normal dispersion regime propagation and are to be compared with the results in Fig. 4 in the main manuscript. 

Although the images in the MNIST Fashion dataset are significantly more complex than the MNIST handwritten digits, we see similar trends in how the performance of the fiber-based ELM depends on different system parameters.  
As expected, given the significantly more complex image structure, the differences seen between the training and test accuracies are greater, and the test accuracies themselves (79-80\%) are lower than obtained for handwritten digits. This reduced accuracy for the MNIST Fashion dataset compared to the MNIST digit dataset is also seen with other machine learning approaches, but the main conclusion here is that even with this significantly more complex data, fiber propagation nonetheless shows improvement compared to linear regression. This result adds to the previous experimental findings that have reported the potential of fiber-based ELMs to be applied to a range of classification tasks  \cite{Fischer-2023,Lee-2024,Hary-2025,Saeed-2025}. 

\vskip 1cm

\begin{figure*}[ht!]
\centering
\includegraphics[width=0.75\linewidth]{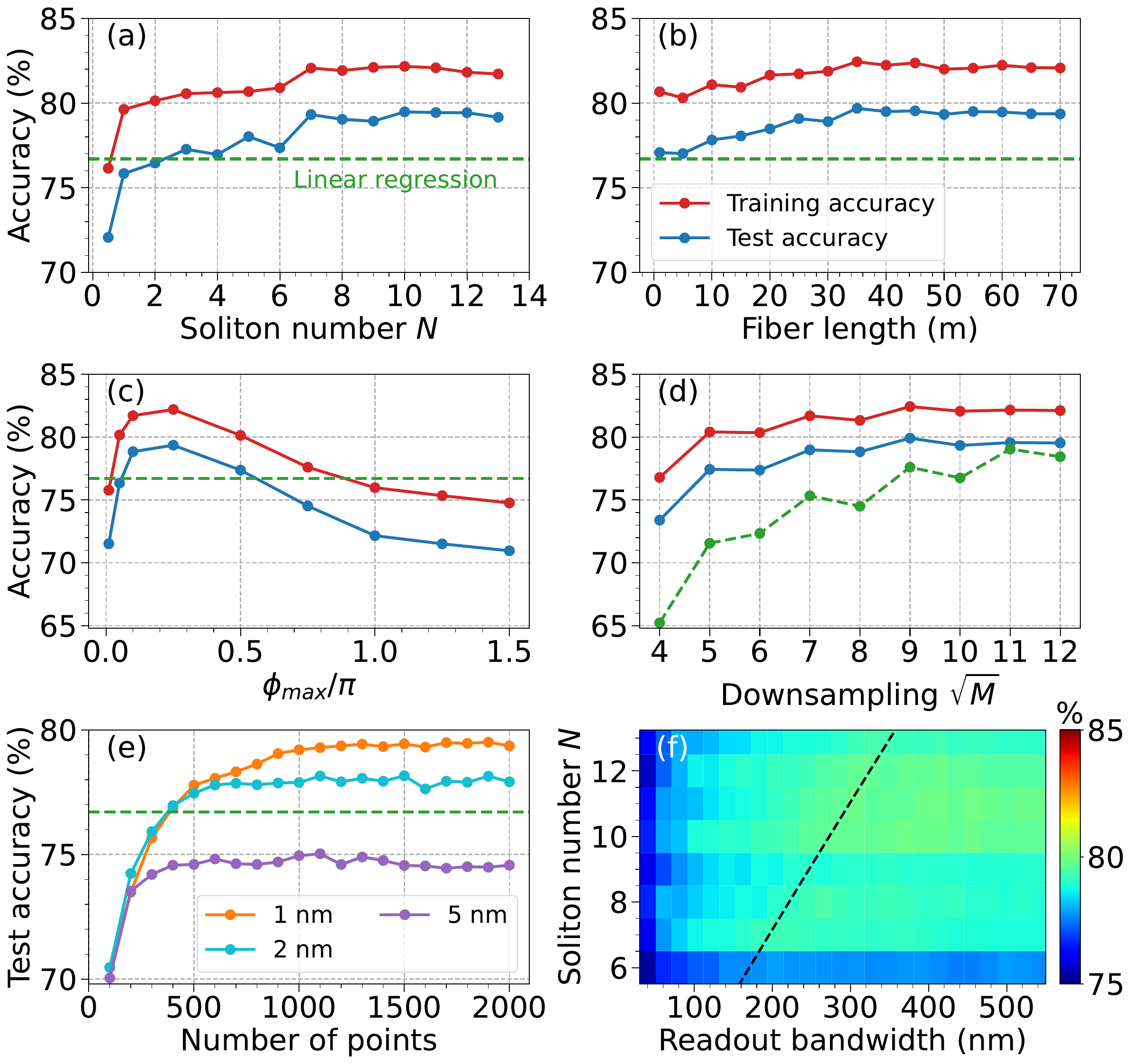}
\caption{MNIST Fashion results - anomalous dispersion.  For $10 \times 10$ downsampling, $\phi_\mathrm{max} = 0.25 \pi$, and 70~m fiber, training (red) and test (blue) accuracy is shown vs: (a) soliton number $N$; (b) fiber length. Green dashed line: linear benchmark. Figures (c-d) use the same color code. For $N = 10$ and 70~m fiber, accuracy is shown: (c) for different $\phi_\mathrm{max}$, and (d) for input downsampling to $\sqrt{M} \times \sqrt{M}$. (e) Test accuracy varying readout bins and convolution bandwidth as indicated. (f) False color plot of test accuracy dependence on $N$ and readout bandwidth (around 1550~nm). Dashed line:  $-20$~dB output  bandwidth.}
\label{fig:ADfashion}
\end{figure*}

\begin{figure*}[ht!]
\centering
\includegraphics[width=0.75\linewidth]{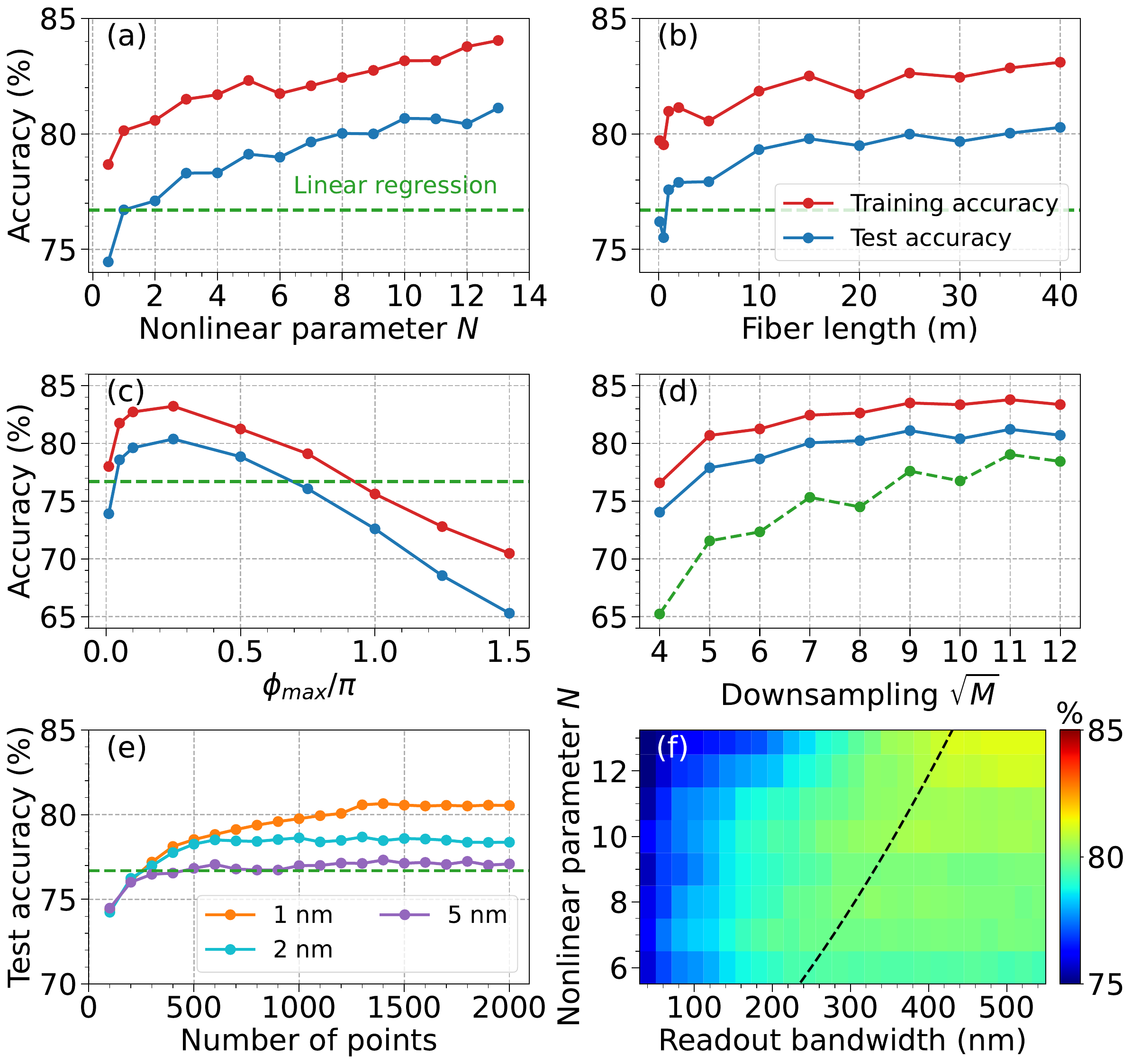}
\caption{MNIST Fashion results - normal dispersion. For $10 \times 10$ downsampling, $\phi_\mathrm{max} = 0.25 \pi$, and 40~m fiber,  training (red) and test (blue) accuracy is shown vs: (a) nonlinear parameter $N$; (b) fiber length. Green dashed line: linear benchmark. Figures (c-d) use the same color code. For $N = 10$ and 40~m fiber, accuracy is shown: (c) for different $\phi_\mathrm{max}$, and (d) for input downsampling to $\sqrt{M} \times \sqrt{M}$. (e) Test accuracy varying readout bins and convolution bandwidth as indicated. (f) False color plot of test accuracy dependence on $N$ and readout bandwidth (around 1550~nm). Dashed line:  $-20$~dB output bandwidth.}
\label{fig:NDfashion}
\end{figure*}

\newpage

\section{Future Research Directions}

Complementing the brief general conclusions in the main manuscript, it is useful to consider future areas of work.  We first remark that the experimental results reported to date are promising, and show that fiber propagation can indeed result in an effective transformation of input-encoded information to a high-dimensional space, suitable for regression and readout in an ELM framework.  

The particular advantage of a simulation model is that it allows us to study and isolate individual effects in a way that is not always possible in experiment, and the particular result here identifying the limiting effect of quantum noise is one such example.  On the other hand, there may be effects encountered in real experiments that are not included in our model. These include notably polarization evolution which could have a significant effect on the results, but which also may be possible to consider as an additional degree of freedom, encoding information on different polarization states in the presence of nonlinear polarization evolution, and then implementing polarization sensitive readout. This is something where a numerical proof-of-principle could be beneficial before implementing experiments, and vector GNLSE models exist that have been well-tested and which are widely available \cite{Agrawal-2019}.  In a similar vein, nonlinear multimode propagation (with spatial encoding and readout) is another obvious area for future study, although numerical computation here can be very demanding as it requires spatio-temporal integration. A fully realistic model may also need to precisely  include the physical process by which input information was encoded on the input pulses. For example, using a grating-based system with a spatial-light-modulator for Fourier-domain encoding, there may be residual spatial overlap between the zero-order and first-order (diffracted) beams, which would require the inclusion of additional noise on the input pulse initial conditions. We note however, that this effect should be able to be mitigated with suitable geometrical optimization and is typically considered negligible with optimized commercial waveshaping device around 1550~nm. 

We also note here that the readout process we study is based on measurements of output spectral intensity, as that is the approach that has been used in experiments to date. This is consistent with the quantum noise model that we use which has been previously shown to yield quantitative agreement with measurements of supercontinuum intensity noise over a range of experimental conditions. However, it is possible to envisage readout based on measurements of spectral phase (using some form of local oscillator as a reference) and in this case it may be necessary to consider more accurate models of input pump noise \cite{Frosz-2010}.  In this context, we also note that phase-readout also opens up the possibility of exploiting the techniques of coherent communications in such a fiber-based ELM platform, potentially greatly increasing the density of data encoding over reduced bandwidths.

% \bibliography{SC_ELM}


%

\end{document}